\documentclass{article}

\usepackage[final]{neurips_2020_ml4ps}

\DeclareUnicodeCharacter{0301}{\'{e}}

\usepackage{natbib}
\usepackage{comment}
\usepackage[utf8]{inputenc} 
\usepackage[T1]{fontenc}    
\usepackage{hyperref}       
\usepackage{url}            
\usepackage{booktabs}       
\usepackage{amsfonts}       
\usepackage{nicefrac}       
\usepackage{microtype}      
\usepackage{graphicx}
\usepackage{textcomp}
\usepackage{gensymb}
\usepackage{subfigure}
\usepackage{wrapfig}
\usepackage[font=small,skip=0pt]{caption}
\vspace{-8mm}
\title{A Deep Learning Approach for Characterizing \\ Major Galaxy Mergers}
\vspace{-8mm}
\author{%
  \textbf{Skanda Koppula${}^1$, Victor Bapst${}^1$, Marc Huertas-Company${}^{2,3,4,5}$} \\
  \textbf{Sam Blackwell${}^1$, Agnieszka Grabska-Barwinska${}^1$, Sander Dieleman${}^1$} \\
  \textbf{Andrea Huber${}^1$, Natasha Antropova${}^1$, Mikolaj Binkowski${}^1$, Hannah Openshaw${}^1$} \\
  \textbf{Adria Recasens${}^1$, Fernando Caro${}^{2}$, Avishai Dekel${}^{6}$, Yohan Dubois${}^{7}$} \\
  \textbf{Jesus Vega Ferrero${}^{8,9}$, David C. Koo${}^{10}$, Joel R. Primack${}^{11}$, Trevor Back${}^1$} \\
  \\
  \vspace{0.5mm}
  ${}^1$DeepMind, London, UK N1C 4AG\\
  ${}^2$LERMA, Observatoire de Paris, PSL Research University, CNRS, \\
  \vspace{0.5mm}
  Sorbonne Universit\'es, UPMC Univ. Paris 06,F-75014 Paris, France\\
  \vspace{0.5mm}
  ${}^3$Univerist\'e de Paris, 5 Rue Thomas Mann - 75013, Paris, France\\
  ${}^4$Departamento de Astrof\'isica, Universidad de La Laguna, \\
  \vspace{0.5mm}
  E-38206 La Laguna, Tenerife, Spain\\
  \vspace{0.5mm}
  ${}^5$Instituto de Astrof\'isica de Canarias, E-38200 La Laguna, Tenerife, Spain\\
  \vspace{0.5mm}
  ${}^6$Racah Institute of Physics, The Hebrew University, Jerusalem 91904 Israel\\
  ${}^7$Institut d'Astrophysique de Paris, Sorbonne Université \\
  \vspace{0.5mm}
  CNRS, UMR 7095, 98 bis bd Arago, 75014 Paris, France \\
  \vspace{0.5mm}
  ${}^{8}$IFCA, Instituto de Física de Cantabria (UC-CSIC), Av. de Los Castros s/n 39005 Santander, Spain,\\
  ${}^{9}$Artificial Intelligence Research Institute (IIIA-CSIC), Campus UAB, Bellaterra, Spain \\
  ${}^{10}$UCO/Lick Observatory, Department of Astronomy and Astrophysics, \\
  \vspace{0.5mm}
  University of California, Santa Cruz, CA 95064, USA \\
  \vspace{0.5mm}
  ${}^{11}$Physics Department, University of California, Santa Cruz, CA 95064, USA \\
}

\begin{document}

\maketitle
\vspace{-8mm}
\begin{abstract}
\vspace{-3mm}
Fine-grained estimation of galaxy merger stages from observations is a key problem useful for  validation of our current theoretical understanding of galaxy formation. To this end, we demonstrate a CNN-based regression model that is able to predict, for the first time, using a single image, the merger stage relative to the first perigee passage with a median error of 38.3 million years (Myrs) over a period of 400 Myrs. This model uses no specific dynamical modeling and learns only from simulated merger events. We show that our model provides reasonable estimates on real observations, approximately matching prior estimates provided by detailed dynamical modeling. We provide a preliminary interpretability analysis of our models, and demonstrate first steps toward calibrated uncertainty estimation.
\end{abstract}
\vspace{-6mm}

\section{Introduction}
\label{submission}
\vspace{-4mm}

Galaxy merging plays a fundamental role in our current theoretical understanding of galaxy formation. Mergers significantly affect galaxy morphology, converting rotationally-supported disk galaxies into velocity-dispersion-supported elliptical galaxies~\citep{1964ApJ...139.1217T}. Gas-rich mergers at high redshift have also been shown to trigger central gas inflow, starburst, and galactic bulge formation~\citep{2015MNRAS.450.2327Z}, and gas-poor mergers enlarge the radii of elliptical galaxies ~\citep{2012ApJ...744...63O}.

Although mergers are present in all theoretical models, observational evidence of their potential effects on galaxies remains elusive. A key challenge in correlating observations with galaxy transformations is calibrating the merger stage: galaxy merging is by definition a dynamical process that takes several million years, and observations that we can record in our lifetime provide only a single time-slice snapshot of such a process. The two approaches used to identify mergers in the sky --- counting pairs of galaxies through deep spectroscopy \citep{2019ApJ...876..110D} and identifying indicative morphological perturbations~\citep{2012ApJ...747...34B} --- both present biases based on the assumed observability of merger stages. Accurate merger stage estimation is also useful for measuring a global \emph{merger rate}: the number of mergers per unit time and volume in the universe. Determining the observability time scale is crucial to measuring these merger rates. This rate is useful for validating cosmological models~\citep{lotz2011major}.

In recent years, there have been several attempts to calibrate galaxy merger detection using state-of-the art simulations that provide dynamical information which is lacking in observations \citep{2008MNRAS.391.1137L,2019MNRAS.486.3702S}. Machine learning has emerged as a strong tool to learn merger properties in simulations \citep{pearson2019identifying, snyder2019machine, ferreira2020galaxy}. These preliminary works have shown that deep learning can successfully classify galaxies into interacting and non-interacting systems using simulation-provided labels. The domain shift to observations still remains a challenge though, as by definition, there is no available ground truth in the observations. Indirect sanity checks such as visual example inspection or comparing with standard morphologies can be undertaken, but it is still difficult to control for all possible systematics~\citep{ferreira2020galaxy, pearson2019identifying}.

In this work, we go several steps further into the characterization of galaxy mergers using deep learning. First, we move from a classification to a regression problem to predict the exact time of a given image of a merger within a merger process. We show that merger stage prediction with a median absolute error of 38.3 million years (Myrs) over a window of 400 Myrs is possible based on a single image and without using any dynamical modeling. Second, we test our model on a well-known system, the Antennae galaxies. Our models trained on simulation snapshots successfully predict the merger stage to the the correct order of magnitude on real observations, matching estimates produced by detailed dynamical modeling. We also explore first steps to measure model uncertainty, and uncover visual indicators on which the model relies.
\vspace{-4mm}
\section{Data}
\vspace{-3mm}
\subsection{Simulation Data}
\vspace{-2mm}
\textbf{2.1.1 Cosmological Simulation:} We use the cosmological hydrodynamical simulation Horizon-AGN ~\citep{2014MNRAS.444.1453D}. The simulation employs an adaptive mesh refinement Eulerian hydrodynamics procedure using RAMSES~\citep{2002A&A...385..337T}. Galaxies are then identified using the AdaptaHOP structure finder~\citep{2004MNRAS.352..376A} over the stellar distribution, using a minimum stellar mass of $10^8$ solar masses. The merger trees for the identified galaxies are then built using the procedure outlined in~\cite{2009A&A...506..647T}. More details on the Horizon-AGN simulation and structure finding are provided in the Appendix.

\textbf{2.1.2 Selection of Galaxy Mergers:} We consider only galaxies more massive than $10^{10}$ solar masses in the redshift range $z=[0.5, 3]$. This is intended to match current deep Hubble Space Telescope observations such as the CANDELS survey~\citep{candels}. We then use the galaxy merger trees from the simulation to select major galaxy mergers following a standard approach of thresholding the stellar mass ratio between the secondary and main progenitor galaxies such that $M_1/M_2<4$ \citep{2015MNRAS.449...49R}.
\label{sec:selection}
 
After this initial selection, we build \emph{merger sequences} which are a complete tracking of the merger process with a time resolution of $\sim 17 $ Myrs. To do so, we first follow each progenitor backward in time, until the progenitors reach a calibrated distance from each other prior to merging (Appendix 5.2). We call $t_{s}$ the time between the sequence's beginning and the first encounter between galaxies ($t_{first\ pass}$). We follow the sequence forward in time by the same amount, so that the duration of the entire merger sequence is $\Delta_T = 2\times t_{s}$. The \emph{i}th snapshot in the sequence occurs at a time $t_i$. Snapshots with time $-t_{s} < t_i < 0$ are considered to be pre-merger, and snapshots with a time $0 < t_i < +t_{s}$ are considered to be post-merger. Since the absolute duration of a merger depends at first order on the dynamical time, we normalize all values by the cosmological dynamical time which we estimate to be $t_{dyn}\sim0.14 t_H$. Here, $t_H$ is the Hubble time $1/H(z)$ at the observation's redshift.
\label{sec:normalization}

\begin{figure}[h]
\hspace{1mm}
 \includegraphics[scale=0.5]{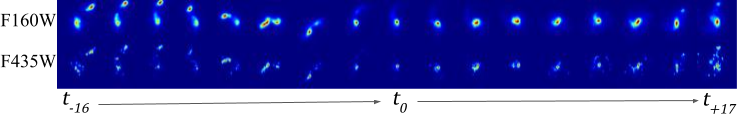}
 \caption{17 samples from a galaxy merge sequence with 34 observations, with an average redshift of 2.146. We show two channels of seven. The first observation ($t_{-16}$) corresponds to a normalized time of -0.24, and the last $t_{+17}$ corresponds to $0.58$. The observation at time $t_{first\ pass}$ is indicated by $t_0=0$.}
 \label{fig:sequence}
 \vspace{-5mm}
\end{figure}

\textbf{2.1.3 Image Generation:} We generate \emph{observed} images of all the snapshots of the selected merger sequences.  Images are produced to replicate properties of Hubble Space Telescope imaging of the CANDELS survey in seven different filters going from the near UV to the near IR (F435W-F160W). We use SUNSET for image generation \citep{ 2017MNRAS.467.4739K, 2019MNRAS.486.5104L}, which models the emission of all galaxy photons to produce an image in the observed-frame.  We generate three different projections along the main axes of the simulations $(X,Y,Z)$. For this work, dust effects in the image generation are not included for computational reasons. We use noiseless images since we want to test whether deep networks can generalize well enough to learn the properties of galaxy mergers based only on examples. We therefore want to maximize the amount of signal in this proof-of-concept work. An example of a merger sequence is shown in Figure~\ref{fig:sequence}.

\subsection{Antennae Observations}
In addition to simulation data, we use archival Hubble Space Telescope observations of the Antennae galaxies (NGC 4038/NGC 4039) to test our model\footnote{\url{https://www.spacetelescope.org/projects/fits_liberator/antennaedata/}}. This well-known system is an archetypal major merger at redshift $z=0.05$ which has been extensively modeled \citep{2008AN....329.1042K,2018MNRAS.475.3934L}. Observations have two channels: F160W and F850LP. In order to better match the properties of the higher redshift training set, we modify the Antennae system observations as if it was observed at high redshift. Given that we are not interested in the absolute flux, we simply apply a spatial scaling to match the angular scale at $z=1.5$. We do not apply dimming to match the high SNR in the training set.
\vspace{-4mm}
\section{Methods}
\vspace{-3mm}
\subsection{Image Pre-processing}
\vspace{-1mm}
Images produced by the Horizon-AGN simulation go through a multi-step pre-processing pipeline. Views from each filter (F160W-F435W) are stacked in the channel dimension of each image. Images are then cropped to a $80 \times 80$ window centered around the point with maximum total intensity. We take this approach to avoid regression target leakage; a simple rescaling would yield an image with apparent galaxy size that correlates with merge state. This resolution-label correlation is not present in real galaxy views, so we take care to avoid label leakage and model degeneracy. Images are augmented before being fed as model input: randomly flipped, rotated in increments of $90 \degree$, jittered, and rescaled. Regression labels are produced through the normalization procedure described in Section~\ref{sec:selection}.

Our Horizon-AGN simulation dataset consists of 6337 galaxy merge sequences, with an average sequence length of 32 time-steps symmetrically straddling $t_0$. Across all sequences, there are 203667 individual observations, with three views per observation. We divide our simulation images into train, validation, and test datasets with a 80\%-10\%-10\% split. Projections from the same merger never occur in more than one split.
\vspace{-2mm}
\subsection{Models and Training Methodology}

We use a convolutional neural network to regress merger time. In particular, we employ a standard ResNet-50 architecture \citep{resnet} to process each input image and produce two regression outputs: the merge time estimate $t$ and an uncertainty score $\sigma$. For models with mass and redshift, we add in a fully-connected layer to embed the mass and redshift features before adding this to each ResNet-block output. We also train models specifically to evaluate Antennae observations, only using the F160W and F850LP channels in our training dataset to match the Antennae samples.

To simultaneously learn uncertainty ($\sigma$) and merger time ($t$), we minimize the sum of a scaled MSE and $\sigma$ estimate, as in \cite{uncertaintyestimation}: $\frac{\log {\sigma}^2}{2}+\frac{(t-\hat{t})^2}{2\sigma^2}$. This is effectively minimizing the log-likelihood criterion for a normal mean/variance estimate. We also add in an 0.0001-weighted L2 regularization term on the trainable weights, yielding our final loss criterion.

We use a standard Momentum SGD optimizer with global norm-based gradient clipping set to 5.0 \citep{momentumsgd}. We set an initial learning rate of 0.025 and use a stepwise-decay schedule, reducing by a factor of 0.1 after 50K and 100K steps. Each model completes training after 150K steps, roughly 7 hours using 2 V100 GPUs.

For our Antennae evaluation, we employ the test-time augmentation methodology proposed in \cite{testtimeaugmentation} to better address the simulated-to-real domain gap. We use their self-consistency loss across augmented image views to fine-tune the model for an additional twenty steps.
\vspace{-3mm}
\section{Results}
\vspace{-2mm}
\subsection{Testing on Simulation}
\vspace{-2mm}
We first evaluate trained models on our test split of simulation images. On the full range of merger sequence snapshot times ($\pm 400$ Myrs), our regression models obtain an root mean square error of 144.1 Myrs. We find that our model's largest errors skew toward the edges of the time interval; the largest errors are early pre-mergers, and the median absolute error is 69.35 Myrs. Interestingly, we find that we obtain slightly better estimation accuracy using a model trained on samples $\pm 200$ Myrs range around the merge. On this time window, the model obtains an RMSE of 68.153 Myrs and median absolute error of 38.391 Myrs. When classifying mergers as either pre- and post-merger (thresholding the predictions and ground truth by $t>0$ or $t\leq0$), our model obtains 86\% classification accuracy on our full range of merger sequences, comparable to prior work~\citep{ferreira2020galaxy}.

\begin{figure}
\vspace{-5mm}
\centering
\includegraphics[width=0.7\textwidth]{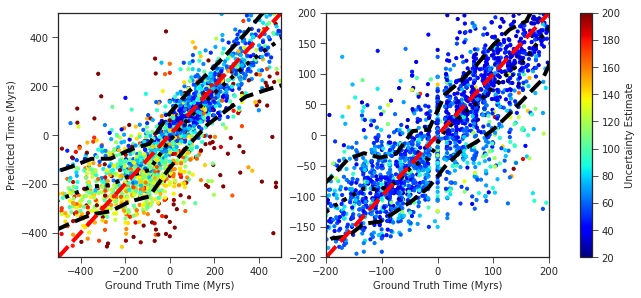}
\caption{Alignment of ground truth and predictions for $\pm 400$ Myrs (left) and $\pm 200$ Myrs (right). Red dashed line: perfect alignment. Black dotted line: the median prediction. Black dashed lines: the prediction variance.}
\label{fig:gtprediction}
\vspace{-5mm}
\end{figure}

Figure~\ref{fig:gtprediction} shows alignment between ground truth and predictions for both these models, along with each model's scaled uncertainty estimate. As previously observed on the full range, our model's median prediction diverges from targets for early pre-mergers. Otherwise, we find reasonable ground truth/prediction alignment, especially for our half-range models. Our uncertainty estimate seems to visually indicate examples outside the scatter (dark brown in the full-range interval, light blue in the half-range interval), and grows more uncertain toward the weakest parts of the alignment.

\vspace{-1mm}
\subsection{Testing on Antennae Observations}

\begin{figure}[h]
\vspace{-2mm}
\centering
\includegraphics[width=0.3\textwidth]{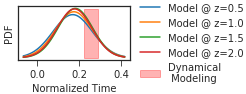}
\hspace{5mm}
\includegraphics[width=0.6\textwidth]{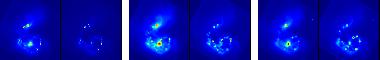}
\vspace{2mm}
\caption{Normalized time predictions for the four redshift variants of the Antennae observation, with prior estimates highlighted in red (left) and the two-filter Antennae observations for $z=0.5, 1.0,$ and $1.5$ (right).}
\label{fig:antennae}
\vspace{-4mm}
\end{figure}

We tested our two-filter, simulation-trained model on four redshift variants of the Hubble space Telescope observations of the Antennae galaxy ($z=0.5, 1.0, 1.5$ and $2.0$). Our model regressed normalized time estimates of $\mu = \{0.167, 0.173, 0.181, 0.177\}$ with a model estimated $\sigma = \{0.094, 0.088, 0.082, 0.083\}$, respectively. Similar regression estimates suggest stability of the model across redshifts. Existing dynamical models of the system~\citep{2008AN....329.1042K,2018MNRAS.475.3934L} estimate that the time of observation is between 500-600 Myrs after first passage. This corresponds to normalized times of $0.24-0.29$ at $z=0.05$ (Figure~\ref{fig:antennae}). Our predictions are therefore in agreement within 1$\sigma$ despite resulting from crude approximations in simulation (lacking dust, noise, etc.). This is extremely encouraging, as this is the first time a model trained on cosmological simulation has been applied to an observed merger snapshot that has independent measurements.

\subsection{Visual Model Analysis}
\begin{wrapfigure}{r}{0.4\textwidth}
\vspace{-14mm}
\centering
\includegraphics[width=0.4\textwidth]{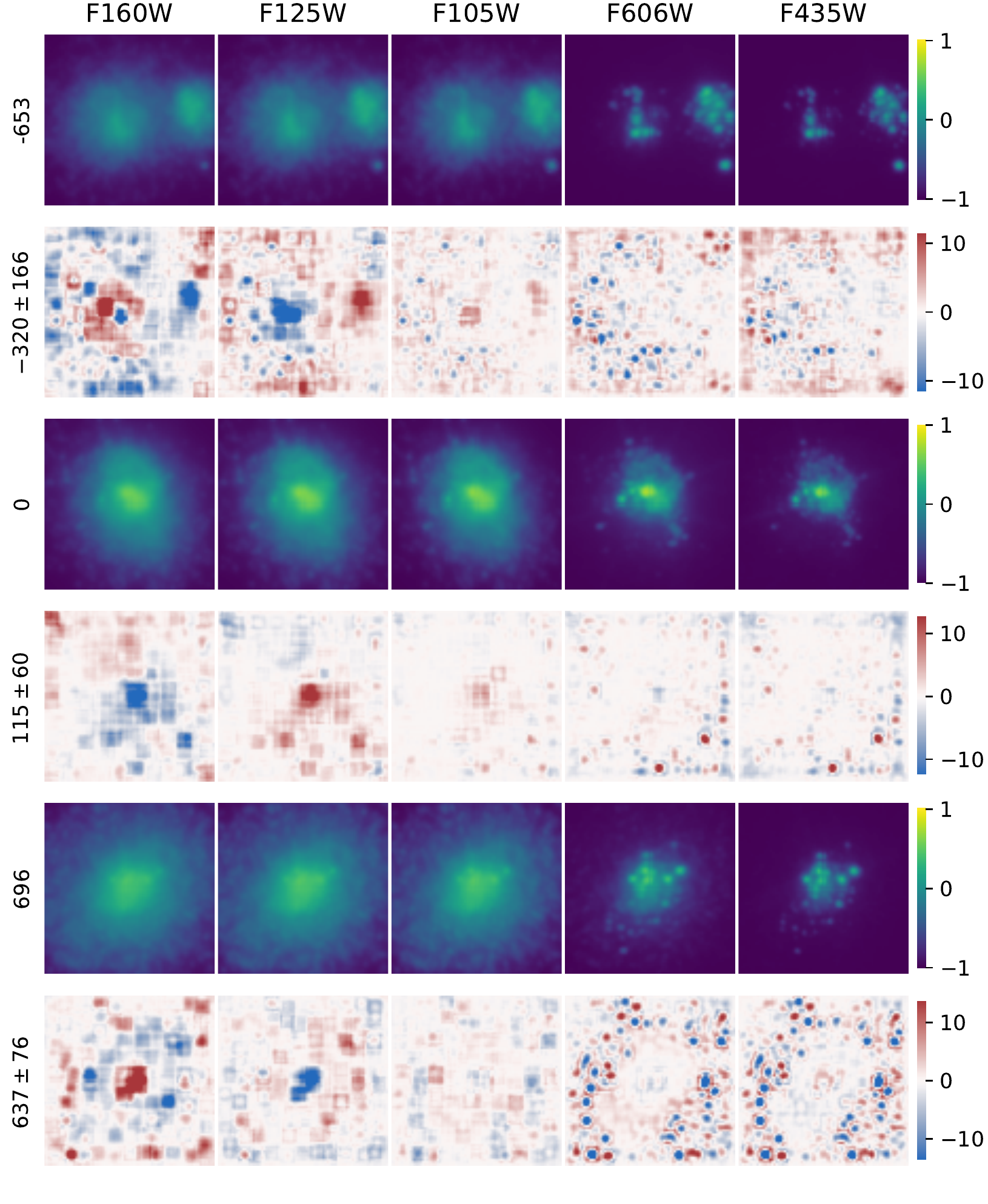}
\caption{Example of input gradients within a merge sequence. Rows show different times in the sequence (image+activation). Columns indicate different filters from near -infrared (F160W) to near-UV (F435W).}
\vspace{-8mm}
\label{fig:gradients}
\end{wrapfigure}

\vspace{-2mm}
We examined regions of input used by the model to produce its predictions. In Figure~\ref{fig:gradients}, we visualise input gradients,  which reflect the  sensitivity of the network prediction to small changes in the input. We show five channels of three snapshots within a merge sequence  (pre-, middle-, and post- from top to bottom) with redshifts of 0.98, 0.84, and 0.70, respectively.

We observe that the patterns of attention depend on wavelength and on the merger stage as expected. Infrared filters (optical rest-frame) tend to focus on the central parts of the galaxies where most of the stellar light comes from. In the pre-merger phase, we clearly see the cores of the two galaxies highlighted. UV rest-frame bands focus more of their attention in the outskirts of the system. This is particularly visible in the post-merger phase. It is likely capturing young stars being formed in the outer regions as a consequence of the interaction.

\vspace{-2mm}
\subsection{Conclusions}
\vspace{-2mm}
Our work shows that temporal constraints on astrophysical observations can be established. We provide evidence that such approaches can be effectively applied to real world observations, while providing some measure of interpretability. We encourage study of the effects of mergers on galaxy evolution from a computational perspective.
\vspace{-6mm}
\section{Appendix}
\vspace{-2mm}
\textbf{5.1. Horizon-AGN and AdaptaHOP Configuration.} \label{appendix:horizonagn} Horizon-AGN was run with a co-moving box size of Lbox = $100 h^{-1}$ Mpc, that contains $1024^3$ DM particles, and that was run considering initial conditions drawn from the WMAP-7 cosmology. The simulation employs the adaptive mesh refinement Eulerian hydrodymamics
code, RAMSES~\citep{2002A&A...385..337T}, and the initially coarse 10243 grid is adaptively
refined, in a quasi-Lagrangian manner, down to a spatial resolution of 1 proper kpc. The AdaptaHOP structure finder~\citep{2004MNRAS.352..376A} was used to identify galaxies, using a minimum threshold of 50 stellar particles (corresponding to a minimum stellar mass of $10^8$ solar masses). The merger trees were built considering 758 time steps that cover a redshift range
spanning from $z=7$ to $z=0$ and with a time difference of $\sim 17$ Myrs on average between two successive time steps.

\textbf{5.2. Galaxy Selection.} \label{appendix:galaxyselection} We look for an increase in the mass of a galaxy due to the contribution of more than one progenitor from the previous time step. If a galaxy has more than one progenitor and the ratio between the stellar mass provided by the secondary and the main progenitor is equal or larger than 1:4, the galaxy is considered as a major merger, and therefore enters our selection. To construct the entire merger sequence from $t_{first\ pass}$, we fellow the progenitors until the secondary progenitor is separated from the main progenitor by a distance larger than four times its effective radius. This is an arbitrary selection that defines the beginning of our merger sequence. It has been calibrated empirically to properly bracket all the different phases of a merger

\textbf{5.3. Image Generation.} For each identified galaxy in the simulation, we define a cubic volume centered around the galaxy with an edge length of eight times the radius of the galaxy (in this case, defined as the average between the three semi-axes obtained when fitting an ellipsoid to the stellar mass distribution of the galaxy). This volume should contain the stellar particles from the main galaxy as well as those from any close companion, in order to capture both galaxies involved in the merger. The stellar particles contained within the volume are used as an input to SUNSET, along with the spectral response of the different filters. SUNSET computes the fluxes corresponding to the inputs using the stellar models of \cite{2003MNRAS.344.1000B} and a \cite{2003PASP..115..763C} IMF.    Finally, the integration of the SED in each pixel and the redshift of the galaxy are used to generate an image in the observed frame.

\textbf{5.4. Broader Impacts} This work proposes an approach for fine-grained estimation of galaxy merger stage using astrophysical simulations. We believe that this work will allow astronomers to improve understanding of galaxy formation by tracking down with unprecedented accuracy the impact of mergers on galaxy transformations over cosmic time. More broadly, this work may be of interest to researchers in computational astronomy and applied machine learning. We believe there is little scope to misuse the artifacts of this work, which uses computational methods to analyze astrophysical data.

\bibliographystyle{plainnat}
\bibliography{neurips_2020}

\end{document}